\documentclass[prd,amsmath,amssymb,showpacs,superscriptaddress,nofootinbib,twocolumn]{revtex4-1}
\usepackage{epsfig,graphics,subfigure,psfrag,amsmath,amssymb}
\usepackage{lineno}
\usepackage{dcolumn}
\usepackage{bm}
\usepackage{color}
\usepackage{overpic}
\usepackage{multirow}
\usepackage[english]{babel}
\addto{\captionsenglish}{%
 
}
\newcommand{\parity}{\mathcal{P}}
\newcommand{\amp}{\mathcal{A}}

\newcommand{\pip}{\pi^{+}}
\newcommand{\pim}{\pi^{-}}
\newcommand{\taup}{\tau^{+}}
\newcommand{\taum}{\tau^{-}}
\newcommand{\hpp}{H^{++}}

\newcommand{\half}{\frac{1}{2}}
\newcommand{\mhalf}{\frac{-1}{2}}

\begin{document}
\normalsize
\parskip=5pt plus 1pt minus 1pt
\title{\boldmath A theorem about two-body decay and its application for a doubly-charged boson $H^{\pm\pm}$ going to $\tau^{\pm}\tau^{\pm}$}
\author{ Li-Gang Xia\\ \it Department of Physics, Tsinghua University, Beijing 100084, People's Republic of China}
\begin{abstract}
In a general decay chain $A\to B_1B_2\to C_1C_2\ldots$, we prove that the angular correlation function $I(\theta_1,\theta_2,\phi_+)$ in the decay of $B_{1,2}$ is irrelevant to the polarization of the mother particle $A$ at production. This guarantees that we can use these angular distributions to determine the spin-parity nature of $A$ without knowing its production details. As an example, we investigate the decay of a potential doubly-charged boson $H^{\pm\pm}$ going to same-sign $\tau$ lepton pair. 
\end{abstract}
\maketitle

\section{introduction}
After the discovery of the higgs boson $h(125)$~\cite{higgs_atlas, higgs_cms}, we are more and more interested in searching for high-mass particles, such as doubly-charged higgs bosons~\cite{hpp_atlas1, hpp_atlas2, hpp_cms}, denoted by $H^{\pm\pm}$. Once we observe any unknown particle, it is crucial to determine its spin-parity ($J^P$) nature to discriminate different theoretic models. A good means is to study the angular distributions in a decay chain where the unknown particle is involved~\cite{nelson1, nelson2, nelson3, spin_qi, spin_mi}. For the Standard Model (SM) higgs, its spin-parity nature can be probed in the decay modes $h(125)\to W^+W^-/ZZ/\taup\taum$~\cite{JP_atlas1,JP_atlas2,JP_atlas3,JP_cms1,JP_cms2}. The validity of this method relies on that the correlation of the decay planes of $W/Z/\tau$ does not depend upon the polarization of $h(125)$ at production. This is proved in a general case in this paper.  As an example, we also investigate the decay $\hpp\to\taup\taup$, where the spin-statistic relation provides more interesting constraints as the final state is two identical fermions. 

\section{Proof of the theorem}
Let us consider a general decay chain $A\to B_1B_2$ with $B_1\to C_1X_1$ and $B_2 \to C_2 X_2$, where $B_1$ and $B_2$ can be different particles and $C_1X_1$ and $C_2X_2$ can be different decay modes even if $B_1$ and $B_2$ are identical particles. Here we prove a theorem, which states that the angular correlation function $I(\theta_1,\theta_2,\phi_+)$ (defined in Eq.~\ref{eq:diff_xsection2}) in the decay of the daughter particles $B_{1,2}$ is independent upon the polarization of the mother particle $A$. Let $\phi_+$ denote the angle between two decay planes $B_i\to C_iX_i$ ($i=1,2$). Therefore, we can measure the $\phi_+$ distribution to determine the spin-parity nature of the mother particle $A$ without knowing its production details~\footnote{After finishing this work, I was informed that the same statement had been verified in Ref.~\cite{nelson1} in the case that $B_{1,2}$ are spin-1 particles and $C_{1,2}$ and $X_{1,2}$ are spin-$\half$ particles. I also admit that it is of no difficulty to generalize it to any allowed spin values for $B$, $C$ and $X$ as shown in this work.}. 

\begin{figure}[htbp] 
\includegraphics[trim=0 70 0 60, clip, width = 0.45\textwidth]{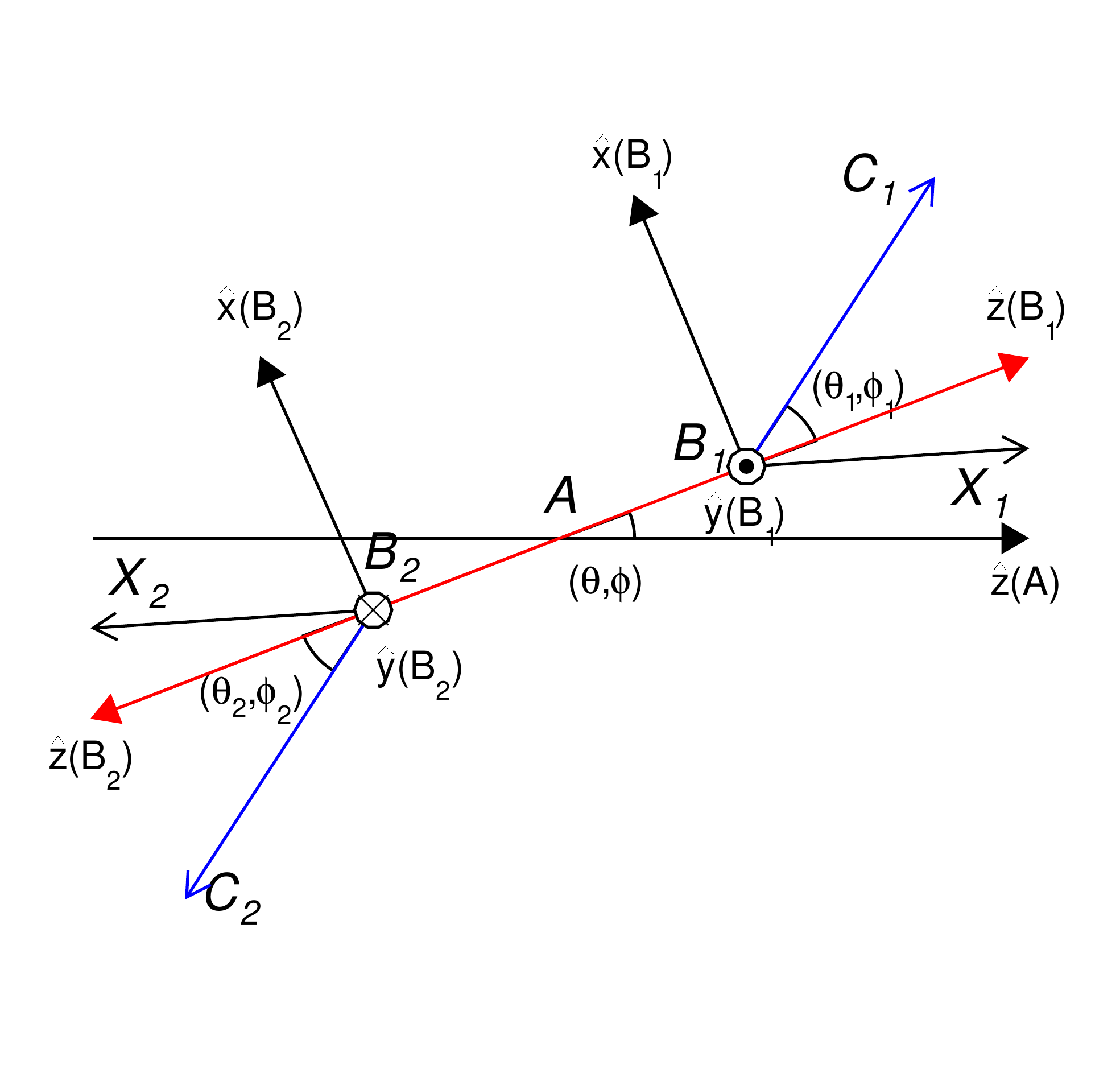} 
\caption{\label{fig:coordinate} The definition of the coordinate system in the decay chain $A\to B_1B_2$ with $B_1\to C_1X_1$ and $B_2\to C_2X_2$. The horizontal arrow represents the flight direction of the mother particle $A$. The red arrows represent the flight directions of $B_{1,2}$ in the rest frame of $A$. The blue arrows represent the flight directions of $C_{1,2}$ in the rest frame of $B_{1,2}$ respectively. $\phi_+$ defined in Eq.~\ref{eq:def_phi_plus} thus represents the angle between the decay plane of $B_1$ and that of $B_2$.}  
\end{figure}

Before calculating the amplitude, we introduce the definition of the coordinate system to describe the decay chain as illustrated in Fig.~\ref{fig:coordinate}. For the decay $A\to B_1B_2$, we take the flight direction of $A$ as the $+z$ axis (if it is still, we take its spin direction as the $+z$ direction), denoted by $\hat{z}(A)$. $\theta$ and $\phi$ are the polar angle and azimuthal angle of $B_1$ in the center-of-mass (c.m.) frame of $A$. For the decay $B_1\to C_1X_1$, we take the flight direction of $B_1$ in the c.m. frame of $A$ as the $+z$ axis, denoted by $\hat{z}(B_1)$ and the direction of $\hat{z}(A)\times \hat{z}(B_1)$ as the $+y$ axis, denoted by $\hat{y}(B_1)$. The $+x$ axis in this decay system is then defined as $\hat{y}(B_1)\times\hat{z}(B_1)$. $\theta_1$ and $\phi_1$ are the polar angle and azimuthal angle of $C_1$ in the c.m. frame of $B_1$. The same set of definitions holds for the decay $B_2\to C_2X_2$. $\phi_+$ is defined in Eq.~\ref{eq:def_phi_plus}. It represents the angle between the two decay planes of $B_i\to C_iX_i$ ($i=1,2$). Here $\phi_1$, $\phi_2$ and $\phi_+$ are constrained in the range $[0,2\pi)$.
\begin{equation}
\label{eq:def_phi_plus}
\phi_+ \equiv \left\{ \begin{array}{ll}\phi_1+\phi_2 \:, &\text{if}\: \phi_1+\phi_2 < 2\pi \\\phi_1+\phi_2-2\pi \:, &\text{if} \: \phi_1+\phi_2 > 2\pi \\
\end{array}\right.
\end{equation}

According to the helicity formalism developed by Jacob and Wick~\cite{JacobWick}, the amplitude is 
\begin{eqnarray}
\amp&& =\sum_{\lambda_1,\lambda_2} F_{\lambda_1\lambda_2}^JD_{M, \lambda_1-\lambda_2}^{J*}(\Omega)\times G_{\rho_1\sigma_1}^{j_1}D_{\lambda_1,\rho_1-\sigma_1}^{j_1*}(\Omega_1)  \nonumber \\
&&\times G_{\rho_2\sigma_2}^{j_2} D_{\lambda_2,\rho_2-\sigma_2}^{j_2*}(\Omega_2) \:. \label{eq:amplitude}
\end{eqnarray}
Here the spin of $A$, $B_1$ and $B_2$ is $J$, $j_1$ and $j_2$ respectively. $M$ is the third spin-component of  $A$. The indices $\lambda_{1,2}$, $ \rho_{1,2}$ and $\sigma_{1,2}$ denote the helicity of $B_{1,2}$, $C_{1,2}$ and $X_{1,2}$ respectively. $D_{mn}^J(\Omega) \equiv D_{mn}^J(\phi, \theta, 0)=e^{-im\phi}d_{mn}^J(\theta)$ and $D_{mn}^J$ ($d_{mn}^J$) is the Wigner $D$ ($d$) function. $F_{\lambda_1\lambda_2}^J$ is the helicity amplitude for $A\to B_1B_2$ and defined  as
\begin{equation} 
F_{\lambda_1\lambda_2}^J \equiv \langle JM;\lambda_1,\lambda_2|\mathcal{M}|JM\rangle \:,
\end{equation}
with $\mathcal{M}$ being the transition matrix derived from  the $S$ matrix. It is worthwhile to note that $F_{\lambda_1\lambda_2}^J$ does not rely on $M$ because $\mathcal{M}$ is rotation-invariant. Similarly, $G_{\rho_i\sigma_i}^{j_i}$ is the  helicity amplitude for $B_i\to C_iX_i$ ($i=1,2$). 

Taking the absolute square of $\amp$ and summing over all possible initial and final states, the differential cross section can be written as  
\begin{eqnarray}
&&\frac{d\sigma}{d\Omega d\Omega_1d\Omega_2} \propto \sum_{M,\lambda_1,\lambda_1^\prime,\lambda_2,\lambda_2^{\prime}}F_{\lambda_1\lambda_2}^JF_{\lambda_1^\prime\lambda_2^\prime}^{J*} e^{i((\lambda_1-\lambda_1^\prime)\phi_1+(\lambda_2-\lambda_2^\prime)\phi_2)} \nonumber \\&&\times d_{M,\lambda_1-\lambda_2}^J(\theta)d_{M,\lambda_1^\prime-\lambda_2^ \prime}^J(\theta) f_{\lambda_1\lambda_1^\prime;\lambda_2\lambda_2^\prime}^{j_1,j_2}(\theta_1,\theta_2) \: , \label{eq:diff_xsection} 
\end{eqnarray}
with 
\begin{eqnarray}
&&f_{\lambda_1\lambda_1^\prime;\lambda_2\lambda_2^\prime}^{j_1,j_2}(\theta_1,\theta_2) \nonumber \\\equiv && \sum_{\rho_1,\sigma_1,\rho_2,\sigma_2} |G_{\rho_1\sigma_1}^{j_1}|^2|G_{\rho_2\sigma_2}
^{j_2}|^2 d_{\lambda_1,\rho_1-\sigma_1}^{j_1}(\theta_1)d_{\lambda_1^\prime,\rho_1-\sigma_1}^{j_1}(\theta_1)\nonumber \\&&\times d_{\lambda_2,\rho_2-\sigma_2}^{j_2}
(\theta_2)d_{\lambda_2^\prime,\rho_2-\sigma_2}^{j_2}(\theta_2) \: .  \label{eq:fBCX}
\end{eqnarray}
Here the summation on $M$ is over the polarization state of $A$ at production. If we do not know the detailed production information, the summation cannot be performed. 

Defining $\delta\lambda^{(\prime)} \equiv \lambda_1^{(\prime)}-\lambda_2^{(\prime)}$, the exponential term in Eq.~\ref{eq:diff_xsection} is equivalent to $e^{i[(\lambda_1-\lambda_1^\prime)\phi_+-(\delta\lambda-\delta\lambda^\prime)\phi_2]}$. Performing the integration on $\phi_2$ and using the definition of $\phi_+$, we have (keeping only the terms related with $\phi_2$) 
\begin{eqnarray}
&&\int_{0}^{2\pi}d\phi_2 e^{i((\lambda_1-\lambda_1^\prime)\phi_1+(\lambda_2-\lambda_2^\prime)\phi_2)} \nonumber \\
=&&\int_{0}^{\phi_+}d\phi_2e^{i[(\lambda_1-\lambda_1^\prime)\phi_+-(\delta\lambda-\delta\lambda^\prime)\phi_2]} \nonumber \\
&&+\int_{\phi_+}^{2\pi}d\phi_2e^{i[(\lambda_1-\lambda_1^\prime)(\phi_++2\pi)-(\delta\lambda-\delta\lambda^\prime)\phi_2]} \:.
\end{eqnarray}
Noting that $(\lambda_1-\lambda_1^\prime)$, $\delta\lambda$ and $\delta\lambda^\prime$ are integers, the integration gives the requirement $ \delta\lambda = \delta\lambda^\prime$. Then the differential cross section in terms of $\lambda_1^{(\prime)}$, $\delta\lambda^{(\prime)}$ and $\phi_+$ is  
\begin{eqnarray}  
&&\sum_{\lambda_1,\lambda_1^\prime,\delta\lambda}F_{\lambda_1,\lambda_1-\delta\lambda}^JF_{\lambda_1^\prime,\lambda_1^\prime-\delta\lambda}^{J*}e^{i(\lambda_1-\lambda_1^\prime)\phi_+} \nonumber \\  
&&\times \sum_{M} d_{M,\delta\lambda}^J(\theta)^2 f_{\lambda_1\lambda_1^\prime;\lambda_1-\delta\lambda,\lambda_1^\prime-\delta\lambda}^{j_1,j_2}(\theta_1,\theta_2) \:. \label{eq:diff_xsection1}  
\end{eqnarray}
According to the orthogonality relations of the Wigner $D$ functions, we obtain
\begin{equation}
\int d_{mn}^J(\theta)^2 d\cos\theta = \frac{2}{2J+1} \:,
\end{equation}
which is independent upon the indices $m, n$. Using this property, we find that integration over $\theta$ of the terms related with $M$ in Eq.~\ref{eq:diff_xsection1} only  provides a constant factor $\sum_{M}\frac{2}{2J+2}$, which is irrelevant to the normalized angular distributions in the $B_{1,2}$ decays. So we finalize the proof of this theorem in Eq.~\ref{eq:diff_xsection2}.
\begin{eqnarray} 
&&I(\theta_1,\theta_2,\phi_+) \equiv \frac{1}{\sigma}\frac{d\sigma}{d\cos\theta_1d\cos\theta_2d\phi_+} \nonumber \\
&& \propto \sum_{\lambda_1,\lambda_1^\prime,\delta\lambda}  
F_{\lambda_1,\lambda_1-\delta\lambda}^JF_{\lambda_1^\prime,\lambda_1^\prime-\delta\lambda}^{J*} \nonumber \\ &&\times e^{i(\lambda_1-\lambda_1^\prime)\phi_+}  
f_{\lambda_1\lambda_1^\prime;\lambda_1-\delta\lambda,\lambda_1^\prime-\delta\lambda}^{j_1,j_2}(\theta_1,\theta_2) \:. \label{eq:diff_xsection2}
\end{eqnarray}

Experimentally, we are interested in the $\phi_+$ distribution, which can be used to measure the spin-parity nature of $A$. We integrate out $\theta_1$ and $\theta_2$ and rewrite $F_{mn}^J\equiv R_{mn}^Je^{i \varphi_{mn}^J}$, where $R_{mn}^J$ and $\varphi_{mn}^J$ are real. The $\phi_+$ distribution turns out to be
\begin{eqnarray}
&&\frac{d\sigma}{\sigma d\phi_+}\propto
\sum_{\lambda_1,\delta\lambda}{R_{\lambda_1,\lambda_1-\delta\lambda}^{J}}^2F_{\lambda_1\lambda_1;\lambda_1-\delta\lambda,\lambda_1-\delta\lambda}^{j_1,j_2} \nonumber \\
&& +\sum_{\lambda_1\neq\lambda_1^\prime}\sum_{\delta\lambda}R_{\lambda_1,\lambda_1-\delta\lambda}^JR_{\lambda_1^\prime,\lambda_1^\prime-\delta\lambda}^{J} F_{\lambda_1\lambda_1^\prime;\lambda_1-\delta\lambda,\lambda_1^\prime-\delta\lambda}^{j_1,j_2} \nonumber \\&&\times\cos[(\lambda_1-\lambda_1^\prime)\phi_++ (\varphi_{\lambda_1,\lambda_1-\delta\lambda}^J-\varphi_{\lambda_1^\prime,\lambda_1^\prime-\delta\lambda}^J)]  \:, ~\label{eq:theorem}
\end{eqnarray}
with
\begin{equation}
F_{\lambda_1\lambda_1^\prime;\lambda_2,\lambda_2^\prime}^{j_1,j_2} \equiv \int f_{\lambda_1\lambda_1^\prime;\lambda_2,\lambda_2^\prime}^{j_1,j_2}(\theta_1,\theta_2) d\cos
\theta_1d\cos\theta_2 \: . \label{eq:FBCX}
\end{equation}
Here the second term in Eq.~\ref{eq:theorem} is obtained using the fact that the summation is invariant with the exchange $\lambda_1 \leftrightarrow \lambda_1^\prime$. 

If the parity is conserved in the decay $A\to B_1B_2$ (namely, $\parity^{-1}\mathcal{M}\parity = \mathcal{M}$ with $\parity$ being the parity operator), we have 
\begin{eqnarray}
&&R_{mn}^J =P_A P_{B_1}P_{B_2}(-1)^{J-j_1-j_2}R_{-m,-n}^J \:, \nonumber \\
&& \varphi_{mn}^J=\varphi_{-m,-n}^J \: , \label{eq:symmetry}
\end{eqnarray}
where $P_{A/B_1/B_2}$ is the parity of $A/B_1/B_2$ and the factor $-1$ is absorbed in $R_{mn}^J$ (namely, we require $0\leq\varphi_{mn}^J<\pi$). Noting that the second summation in Eq.~\ref{eq:theorem} is invariant with the index exchange $(\lambda_1,\lambda_1',\delta\lambda) \leftrightarrow (-\lambda_1, -\lambda_1', -\delta\lambda)$, thus we have
\begin{equation}
\sum_{\lambda_1\neq\lambda_1^\prime}\sum_{\delta\lambda} \cdots = \frac{1}{2} \sum_{\lambda_1\neq\lambda_1^\prime}\sum_{\delta\lambda}\cdots + \frac{1}{2} \sum_{-\lambda_1\neq-\lambda_1^\prime}\sum_{-\delta\lambda}\cdots \: .
\end{equation}
Using the symmetry relation in Eq.~\ref{eq:symmetry}, this summation turns out to be 
\begin{eqnarray}\label{eq:Psum}
&&\frac{1}{2}\sum_{\lambda_1\neq\lambda_1^\prime}\sum_{\delta\lambda}R_{\lambda_1,\lambda_1-\delta\lambda}^JR_{\lambda_1^\prime,\lambda_1^\prime-\delta\lambda}^{J} \times \left\{F_{\lambda_1\lambda_1^\prime;\lambda_1-\delta\lambda,\lambda_1^\prime-\delta\lambda}^{j_1,j_2}\right. \nonumber \\
&&\times \cos[(\lambda_1-\lambda_1^\prime)\phi_++ (\varphi_{\lambda_1,\lambda_1-\delta\lambda}^J-\varphi_{\lambda_1^\prime,\lambda_1^\prime-\delta\lambda}^J)]  \nonumber \\
&&+ F_{-\lambda_1,-\lambda_1^\prime;-\lambda_1+\delta\lambda,-\lambda_1^\prime+\delta\lambda}^{j_1,j_2} \nonumber \\
&&\left.\times \cos[(\lambda_1-\lambda_1^\prime)\phi_+- (\varphi_{\lambda_1,\lambda_1-\delta\lambda}^J-\varphi_{\lambda_1^\prime,\lambda_1^\prime-\delta\lambda}^J)]  \right\} \:.
\end{eqnarray}
Focusing on the expressions of Eq.~\ref{eq:FBCX} and Eq.~\ref{eq:fBCX}, we are able to show that
\begin{equation}\label{eq:PFBCX}
F_{\lambda_1\lambda_1^\prime;\lambda_1-\delta\lambda,\lambda_1^\prime-\delta\lambda}^{j_1,j_2} = F_{-\lambda_1,-\lambda_1^\prime;-\lambda_1+\delta\lambda,-\lambda_1^\prime+\delta\lambda}^{j_1,j_2} \: ,
\end{equation}
using the following property of the Wigner $d$ function 
\begin{equation}
d_{mn}^j(\pi-\theta) = (-1)^{j-n} d_{-m,n}^j(\theta) \: .
\end{equation}
With Eq.~\ref{eq:Psum} and Eq.~\ref{eq:PFBCX}, Eq.~\ref{eq:theorem} can be simplified as 
\begin{eqnarray}
&&\frac{d\sigma}{\sigma d\phi_+}\propto\sum_{\lambda_1,\delta\lambda}{R_{\lambda_1,\lambda_1-\delta\lambda}^{J}}^2F_{\lambda_1\lambda_1;\lambda_1-\delta \lambda,\lambda_1-\delta\lambda}^{j_1j_2} \nonumber \\
&& +\sum_{\lambda_1\neq\lambda_1^\prime}\sum_{\delta\lambda}R_{\lambda_1,\lambda_1-\delta\lambda} ^JR_{\lambda_1^\prime,\lambda_1^\prime-\delta\lambda}^{J}F_{\lambda_1\lambda_1^\prime;\lambda_1-\delta\lambda,\lambda_1^\prime-\delta\lambda}^{j_1j_2} \nonumber \\
&&\times\cos(\varphi_{\lambda_1,\lambda_1-\delta\lambda}^J-\varphi_{\lambda_1^\prime,\lambda_1^\prime-\delta\lambda}^J)\cos[(\lambda_1-\lambda_1^  \prime)\phi_+] 
\:, ~\label{eq:theorem1} 
\end{eqnarray}
This expression is actually the Fourier series for a $2\pi$-periodic even function. Comparing Eq.~\ref{eq:theorem} and Eq.~\ref{eq:theorem1}, we can see that the terms which are odd with respective to $\phi_+$ are forbidden due to parity conservation in the decay $A\to B_1B_2$.


Now we consider the special case that $B_1$ and $B_2$ are identical particles and $B_{1,2}$ decay to the same final state, for example, we will study a doubly charged boson decay $H^{++}\to\taup\taup\to\pip\pip\bar{\nu}_\tau\bar{\nu}_\tau$. For identical particles, the state with the spin $J$ and the third component $M$ is 
\begin{equation}
|JM;\lambda_1\lambda_2\rangle_S = |JM;\lambda_1\lambda_2\rangle +(-1)^{J} |JM;\lambda_2\lambda_1\rangle \: ,
\end{equation}
which satisfies the spin-statistics relation. Here the normalization factor is omitted. The helicity amplitude $F_{\lambda_1\lambda_2}^J = _S\langle JM;\lambda_1\lambda_2|\mathcal{M}|JM\rangle$ has the symmetry$F_{\lambda_1\lambda_2}^J = (-1)^JF_{\lambda_2\lambda_1}^J$. This symmetry relation will further constrain the helicity states, namely, the indices $\lambda_1$, $\lambda_1^\prime$ and $\delta\lambda$ in the summation in Eq.~\ref{eq:diff_xsection2},~\ref{eq:theorem} and ~\ref{eq:theorem1}.

\section{Study of $\hpp\to\taup\taup\to\pip\pip\bar{\nu}_\tau\bar{\nu}_\tau$}
Ref.~\cite{ZpZZ} is an example of the application of this theorem. It studies the decay $Z^\prime \to ZZ \to l^+l^-l^+l^-$, where $B_{1,2}$ are identical bosons. Here we consider the decay chain $\hpp\to\taup\taup\to\pip\pip\bar{\nu}_\tau\bar{\nu}_\tau$. For two spin-$\half$ identical fermions, we write down all states explicitly. The helicity index $\lambda=+\half $ ($\mhalf$) is denoted by $R$  ($L$). 
\begin{eqnarray}|JM;LL\rangle_S =&& (1+(-1)^J)|JM;LL\rangle \nonumber \\&& \parity|JM;LL\rangle_S = -|JM;RR\rangle_S \\|JM;RR
\rangle_S =&& (1+(-1)^J)|JM;RR\rangle  \nonumber \\&& \parity|JM;RR\rangle_S = -|JM;LL\rangle_S \\|JM;LR\rangle_S =&& |JM;LR\rangle+(-1)^J|JM;RL\rangle \nonumber \\&& 
\parity|JM;LR\rangle_S = -|JM;LR\rangle_S   
\end{eqnarray}

The third state is already a parity eigenstate. The first two states can be combined to have a definite parity.
\begin{equation}
(1+(-1)^J)(|JM;LL\rangle\pm|JM;RR\rangle) \:,  \quad P = \mp 1
\end{equation}
In addition, the angular momentum conservation requires $|\lambda_1- \lambda_2|\leq J$. Now we can give the selection rules, which are summarized in  Table~\ref{tab:selection_rules}. We can see that the states with odd spin and even parity are forbidden. For comparison, the selection rules for a neutral particle decaying to spin-$\half$  fermion anti-fermion pair are summarized in Table~\ref{tab:selection_rules1}. 

In future electron-electron colliders, $H^{--}$ may be produced in the process $e^-e^- \to H^{--}$. However,  the reaction rate for a spin-1 $H^{--}$ will be highly suppressed because the vector coupling requires that both electrons have the same handness while the only allowed state is $|LR  \rangle-|RL\rangle$. Similarly, the production rate for a scalar $H^{--}$ is also highly suppressed. This is called ``helicity suppression''.

\begin{table}[htbp]\caption{\label{tab:selection_rules}Selection rules for a particle decaying to two 
spin-$\half$ identical fermions.}
\begin{center}
\begin{ruledtabular}
\begin{tabular}{l l l l }Parity & $J=0$ & $J=2,4,6,\ldots$ & $J=1,3,5,\ldots$ \\
\hline even & $|LL\rangle-|RR\rangle$ &$|LL\rangle-|RR\rangle$ & forbidden \\
\hline \multirow{2}{*}{odd} & \multirow{2}{*}{$|LL\rangle+|RR\rangle$}  &$|LL\rangle+|RR\rangle$ & \multirow{2}{*}{$|LR\rangle-|RL\rangle$} \\
& & $|LR\rangle+|RL\rangle$& \\
\end{tabular}
\end{ruledtabular}  
\end{center}
\end{table}
\begin{table}[htbp]\caption{\label{tab:selection_rules1}Selection rules for a particle decaying to spin-$\half$ fermion 
anti-fermion pair.}
\begin{center}
\begin{ruledtabular}
\begin{tabular}{l l l l}Parity & $J=0$ & $J=2,4,6,\ldots$ & 
$J=1,3,5,\ldots$ \\
\hline\multirow{2}{*}{even} & \multirow{2}{*}{$|LL\rangle+|RR\rangle$} &$|LL\rangle+|RR\rangle$ & $|LL\rangle-| RR\rangle$ \\
&& $|LR\rangle+|RL\rangle$&$|LR\rangle-|RL\rangle$ \\
\hline  \multirow{2}{*}{odd} & \multirow{2}{*}{$|LL\rangle-|RR\rangle$} &$|LL\rangle-|RR\rangle$ & $|LL\rangle+|RR\rangle$ \\&& $|LR\rangle-|RL\rangle$&$|LR\rangle+|RL\rangle$ \\
\end{tabular}
\end{ruledtabular}
\end{center}
\end{table}

\begin{figure}[htbp] 
\includegraphics[width = 0.45\textwidth]{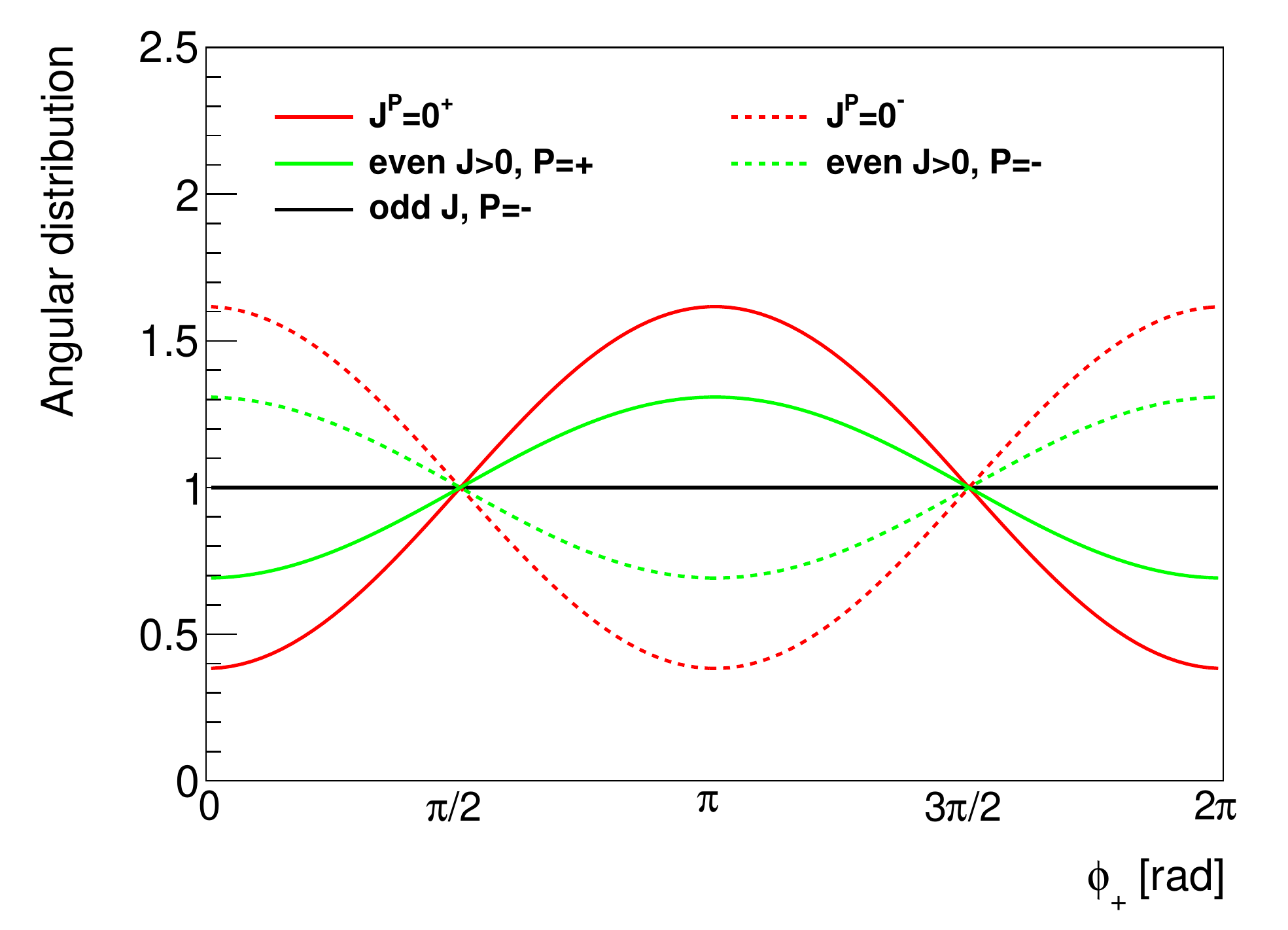} 
\caption{\label{fig:phi_plus} The $\phi_+$ distributions for different $J^{P}$s. The black  line represents odd $J$. The red solid (dashed) curve represents $J^P=0^+ (0^-)$. The green solid (dashed) curve represents even $J>0$ with even (odd) parity assuming $a_J=1$.}  
\end{figure}

Replacing $A$, $B_{1,2}$ and $C_{1,2}$ by $\hpp$, $\taup$ and $\pip$ respectively in Eq.~\ref{eq:amplitude}, the amplitude is 
\begin{eqnarray}
\amp = &&G_{0\half}^\half G_{0\half}^\half e^{iM\phi}\left[F_{RR}^Jd_{M0}^J(\theta)e^{i(\half\phi_1+ \half\phi_2)}\sin\frac{\theta_1}{2}\sin\frac{\theta_2}{2} \right. \nonumber \\ 
&&+\left.F_{LL}^Jd_{M0}^J(\theta)e^{-i(\half\phi_1+\half \phi_2)}\cos\frac{\theta_1}{2}\cos\frac{\theta_2}{2} \right. \nonumber \\
&&-\left.F_{LR}^Jd_{M,-1}^J(\theta)e^{i(-\half\phi_1+\half \phi_2)}\cos\frac{\theta_1}{2}\sin\frac{\theta_2}{2} \right. \nonumber \\
&&-\left.F_{RL}^Jd_{M,1}^J(\theta)e^{i(\half\phi_1-\half\phi_2)} \sin\frac{\theta_1}{2}\cos\frac{\theta_2}{2}\right] \: .\label{eq:amp0} 
\end{eqnarray}
 Here we have only one decay helicity amplitude, $G_{0\half}^\half$, for the $\taup$ decay. This is because $\pip$ is a pseudo-scalar and $\bar{\nu}_\tau$ is right-handed.
 The angular correlation function is  
\begin{eqnarray}
I(\theta_1,\theta_2,\phi_+)  \propto && 1+ \cos \theta_1\cos\theta_2 \:, \quad \text{for odd } J \nonumber \\
I(\theta_1,\theta_2,\phi_+) \propto && 1+a_J^2 + (1-a_J^2)\cos\theta_1\cos\theta_2 \nonumber \\
&& -P_H\sin\theta_1\sin\theta_2\cos\phi_+ \:, \quad \text{for even } J  \quad \quad \label{eq:diff_xsection_oddevenJ}   
\end{eqnarray}
Here for even $J$, $a_J$ is defined as $a_J\equiv|F_{LR}^J|/|F_{RR}^J|$. $P_H$ is the parity of $\hpp$. We can see that the polarization information of $\hpp$ does not appear in the angular distributions. The $\phi_+$ distribution is   
\begin{equation}\label{eq:dist_phi_plus}  
\frac{d\sigma}{\sigma d\phi_+} \propto \left\{\begin{array}{ll} 1 &\text{for odd } J \nonumber \\
1 -P_H\frac{\pi^2}{16}\frac{1}{1+a_J^2} \cos\phi_+ &\text{for even } J \\
\end{array}\right. \: .
\end{equation}
 The $\phi_+$ distributions for different $J^P$s are shown in Fig.~\ref{fig:phi_plus}, where $a_J=1$ is assumed for illustration.  

Here are a few conclusions.
 \begin{enumerate}  
 \item The $\phi_+$ distribution is uniform for odd $J$. 
 \item For $J=0$, the 
helicity amplitudes $F_{LR}^J$ and  $F_{RL}^J$ are forbidden due to angular momentum conservation. Thus $a_J=0$ and the $\phi_+$ distribution becomes
 \begin{equation}
  \frac{d\sigma}{\sigma d\phi_+} \propto 1-P_H\frac{\pi^2}{16}\cos\phi_+ \:,
 \end{equation}
  which is the same as that in the decay $h(125)\to \tau^+\tau^-\to\pip\pim\nu_{\tau}\bar{\nu}
_{\tau}$.  
\item For nonzero even $J$, the $  \phi_+$ distribution depends upon $J$ through the amplitude ratio $a_J$.
 \end{enumerate} 
 
 Experimentally,  it is difficult to reconstruct the $\tau$ lepton information due to the invisible neutrinos~\cite{MMC,mtt_xlg}. But we are able to obtain the decay plane angle $\phi_+$ in some ways (see a  most recent review Ref.~\cite{BBK} and references therein). The so-called impact parameter method~\cite{ImpactParameter} is suitable for the decay $\taup\to\pip\bar{\nu}_\tau$ studied here. It requires that final $\pip$s have significant impact parameters, which condition can be satisfied at high-energy colliders such as the Large Hadron Collider (LHC). 
 
 \section{Conclusions}
 In summary, for  a general decay chain $A\to B_1B_2\to C_1C_2\ldots$, we have proved that the angular correlation function $I(\theta_1,\theta_2,\phi_+)$ in the decay of the daughter particles $B_{1,2}$ is independent upon the polarization of the mother particle $A$ at production. It guarantees that the spin-parity nature of the mother particle  $A$ can be determined by measuring the angular correlation of the two decay planes $B_{i}\to C_i\ldots$ ($i=1,2$) without knowing its production details. This theorem has a simple form if the parity is conserved in the decay $A\to B_1B_2$. Taking a potential doubly-charged particle decay $\hpp\to\taup\taup$ as example, we present the selection rules for various spin-parity combinations. It is found that this decay is forbidden for the $\hpp$ with odd spin and even parity. Furthermore, we show that the angle between the two $\tau$  decay plans is an effective observable to determine the spin-parity nature of $\hpp$. 
 
 \section{Acknowledgement}
 Li-Gang Xia would like to thank Fang Dai for many helpful discussions. The author is also indebted to Yuan-Ning Gao for enlightening discussions. This work is supported by the General Financial Grant from the China Postdoctoral Science Foundation (Grant No. 2015M581062).   
 
\end{document}